\documentclass[10pt]{article}

\usepackage{fullpage}
\usepackage{setspace}
\usepackage{parskip}
\usepackage{titlesec}
\usepackage[section]{placeins}
\usepackage{xcolor}
\usepackage{breakcites}
\usepackage{lineno}
\usepackage{hyphenat}

\usepackage{comment}

\PassOptionsToPackage{hyphens}{url}
\usepackage[colorlinks = true,
            linkcolor = blue,
            urlcolor  = blue,
            citecolor = blue,
            anchorcolor = blue]{hyperref}
\usepackage{etoolbox}

\usepackage{natbib}

\renewenvironment{abstract}
  {{\bfseries\noindent{\abstractname}\par\nobreak}\footnotesize}
  {\bigskip}

\titlespacing{\section}{0pt}{*3}{*1}
\titlespacing{\subsection}{0pt}{*2}{*0.5}
\titlespacing{\subsubsection}{0pt}{*1.5}{0pt}

\usepackage{authblk}
\usepackage[disable]{todonotes}

\usepackage{graphics}
\usepackage{subcaption}
\usepackage[space]{grffile}
\usepackage{latexsym}
\usepackage{textcomp}
\usepackage{longtable}
\usepackage{tabulary}
\usepackage{booktabs,array,multirow}
\usepackage{amsfonts,amsmath,amssymb}
\providecommand\citet{\cite}
\providecommand\citep{\cite}

\newif\iflatexml\latexmlfalse
\usepackage[utf8]{inputenc}
\usepackage[english]{babel}
\usepackage{graphicx}
\usepackage{float}

\begin{document}

\title{Open Source Software Field Research: Spanning Social and Practice Networks for Re-Entering the Field}

\author[1]{Sean P. Goggins}
\affil[1]{University of Missouri}
 %ORCID: 0000-0002-4331-147X

\author[2]{Kevin Lumbard}
\affil{Creighton University}
%\orcid{0000-0001-9306-3040}

\author{Matt Germonprez}
\affil[3]{University of Nebraska at Omaha}
%\orcid{0000-0003-2326-5901}

\author{Caifan Du}
\affil[3]{University of Texas - Austin}
%\orcid{https://orcid.org/0000-0003-2538-607X}

\author{Karthik Ram}
\affil[4]{University of California, Berkeley}
%\orcid{0000-0002-0233-1757}

\author{James Howison}
\affil[3]{University of Texas - Austin}
%\orcid{0000-0002-5702-149X}

\vspace{-1em}

\begingroup
\let\center\flushleft
\let\endcenter\endflushleft
\maketitle
\endgroup

\selectlanguage{english}
\begin{abstract}
Sociotechnical research increasingly includes the social sub-networks that emerge from large-scale sociotechnical infrastructure, including the infrastructure for building open source software. This paper addresses these numerous sub-networks as advantageous for researchers. It provides a methodological synthesis focusing on how researchers can best span adjacent social sub-networks during engaged field research. Specifically, we describe practices and artifacts that aid movement from one social subsystem within a more extensive technical infrastructure to another. To surface the importance of spanning sub-networks, we incorporate a discussion of social capital and the role of technical infrastructure in its development for sociotechnical researchers. We then characterize a five-step process for spanning social sub-networks during engaged field research: commitment, context mapping, jargon competence, returning value, and bridging. We then present our experience studying corporate open source software projects and the role of that experience in accelerating our work in open source scientific software research as described through the lens of bridging social capital. Based on our analysis, we offer recommendations for engaging in fieldwork in adjacent social sub-networks that share a technical context and discussion of how the relationship between social and technically acquired social capital is a missing but critical methodological dimension for research on large-scale sociotechnical research. 
\end{abstract}

\section{Introduction}
The domains of study in social computing, including open source software, social media, crowdsourcing, and citizen science, are increasingly centered on large-scale platforms with vague boundaries between the associated sub-groups of people and their constituent sub-networks. Understanding how and to what extent knowledge transfers between sub-networks in large domains occur is critical for understanding how information, knowledge, and practices are shared between different forms of technologically mediated engagement \citep{introne_advice_2019}. Social computing research is often centered on a single technology and bounded by a particular study's social networks and sub-networks. To advance the field, it is necessary to understand how social capital developed in one socio-technical sub-network can be leveraged as an entryway for scholarship with broader academic and applied outcomes.

Boundaries between sub-networks within open source software, for example, are defined by practices of technology use \citep{harrison_places:_2008} and the movement of knowledge between sub-networks \citep{introne_advice_2019}. Yet, there are limited examples of social computing research that span between social computing sub-networks, with notable exceptions in crisis informatics \citep{hughes_twitter_2009, cipalen_citizen_2007, cipalen_crisis_2009, cipalen_vision_2010, cisarcevic_beacons_2012, cistarbird_chatter_2010, cistarbird_how_2012, cistarbird_voluntweeters:_2011, sutton_backchannels_2008, civieweg_microblogging_2010}. In response, we argue in this paper that engaged field research along with the analysis of technical artifacts can improve how researchers can conduct research across sub-networks boundaries \citep{introne_design_2012, introne_advice_2019, introne_designing_2020, introne_sociotechnical_2016, introne_taming_2015, introne_tracing_2012, goggins_group_2013, goggins_connecting_2014, howison2016software}. In this paper, we contribute a case of sub-network spanning research as an exemplary process for discovering and navigating research across sub-network boundaries in large-scale sociotechnical systems in open-source software. We illustrate and argue for the unattended role of social capital as an essential dimension of engaged field research that incorporates social computing artifacts and data. Our contribution draws on our experiences entering the field \citep{chughtai_entering_2017} as we cross boundaries to study and engage with different sub-networks working on open source software. 

\citet{star_power_2007} identifies the boundary-crossing experience for sociotechnical researchers as a choice between being part of a network (singular) or part of networks (plural). In many cases, field engagement takes on significant long-term investment as researchers simultaneously engage deeply with multiple networks. These time commitments draw scholars away from their primary network of practice in sociotechnical systems research. Our academic communities impart a kind of invisible work regarding the standards for accepting work. When working across networks, the likelihood of experiencing marginality grows. This experience is especially true if a few scholars build bridges between any two scholarly or practice fields. The gap Star and others \citep{asdal_technoscience_2007} identify for sociotechnical researchers remains unresolved more than a decade later. 

The field of practice our research work centers on is open source software (OSS), which has seen tremendous growth in the last decade. The landscape of OSS is vast, and research access to one corner of this landscape no longer necessarily includes access to another corner~\citep{germonprez_eight_2018}. Here, Star's tension between networks \citep{star_power_2007} and the resulting marginalization for the scholar becomes apparent through the common conceptualization of OSS as a single context. This oversimplification of open source software in sociotechnical and social computing discourse exists in HCI's larger tradition of studying complex sociotechnical phenomena from the perspective of being "in here" (in our scholarly network) and, simultaneously "out there" (outside our scholarly network) \citep{taylor_out_2011}. Regardless, engagement in outside networks is called for as researchers build research programs that account for the practices and constructs across distinct sub-networks.

OSS scholarship has a long arc of influence on culture \citep{kelty_two_2008}, often at the intersection between the ideals of open culture and its reality \citep{coleman_coding_2013}. \citet{kelty_two_2008} describes, with depth, the state of those intersections in the middle of the 2000s, and  \citet{coleman_coding_2013} takes a deep dive into the practices and ideals of the Debian (A Linux Kernel "flavor") community during the same period. \citet{dunbar-hester_hacking_2020} provides more contemporary observations centering on open source software as a mechanism for increasing diversity and addressing problems of global political inequality through new technology, pointing out that new technologies often promise liberation but seldom advance the cause. Stemming from prior open source work, we know that OSS does not exist as a single network but exists uniquely in corporate, scientific, public health, academic, and research OSS sub-networks. Techniques for sharing knowledge between sub-networks is critically important for all. Our paper proposes methods for boundary spanning across open source sub-networks to understand OSS better and put into practice how we express OSS as a multi-networked system. 

This paper argues that each corner of OSS is a distinct study sub-network. Each corner of OSS may require a long-term investment by default because the sociotechnical networks are entirely different, something obscured because much technology and underlying philosophy remain the same. In this paper, we frame the social capital developed through long-standing engagement with one sub-network of OSS, corporate open source, as having technical and practical utility to contributors in another sub-network of OSS, open source scientific software \citep{star_ethnography_1999}. We propose that insights gained, instruments defined, and tools built to garner an understanding of one OSS sub-network may afford a researcher some social capital within an adjacent OSS sub-network, as our case describes. The effects can be shorter periods of building relationships and conducting research in a bordering network. However, we anticipate that efforts in our prior contexts do not give us \textit{carte blanche} access to new sub-networks ~\citep{tajfel_social_1982}. As we enter a new network, we return to \citet{star_power_2007}, who observed tensions within our expanding field of view. 

We propose that OSS research can benefit from the development of social capital~\citep{coleman_social_1988} within different OSS sub-networks (corporate, scientific, and others) to have the access necessary to contribute to our scholarly outputs. The following sections describe prior work illustrating the role of two types of social capital and explain how each is conceptually vital for engaged field research in our context of open source software. We then demonstrate how prior engaged field research periodically alludes to social capital as a methodological concern and pull those discrete insights together, much like Agre consolidated insights about the missing social dimension in long-standing political theory \citep{agre1997reinventing, agre_practical_2004}. We then ground our argument that social capital may be missing in boundary-spanning research by describing our reentry into the field, as experienced in our move from corporate to scientific open source software. We conclude with brief recommendations for institutionalizing multi-disciplinary research into the sub-networks of open source software; which from a distance may appear as a singular, monolithic field of sociotechnical materialities \citep{leonardi_materiality_2012} that can be studied without considering their social, contextual, and domain-specific elements. 

\section{Prior Literature}

Social capital applies methodologically, much like moving from one home in one city to a new home in a new town; the transition is never smooth. When a person moves, they must make new friends and rearrange furniture often. The function of the kitchen reveals itself slowly over time, and it takes a year to learn when and where the sun shines in a garden. For large-scale sociotechnical research, the same fundamental skills and essential infrastructure \citep{star_steps_1996, bateson_steps_2000} apply (i.e., lounging, cooking, and gardening), but specifying how those skills and objects are enacted and used in different homes and different neighborhoods is a metaphor for how researchers might approach sub-network spanning fieldwork. Everything looks the same if observed from afar, yet is experienced differently by the people involved depending on the context. Moving requires time to identify differences and similarities that help people adjust and earn the social capital necessary to, for example, borrow a neighbor's shovel. Accruing social capital can also be a cornerstone of engaged fieldwork. It takes time to gather and is easy to lose, but it can be leveraged into research productivity. 

It is safe to say that the researcher accrues social capital in a new sociotechnical context by contributing to the members of the context they are working with---not in the form of ideas or suggestions but documented and applied contributions. It is like how, in the moving metaphor, shoveling our neighbor's walk helps to build the social capital we need to get a "yes" when we ask to borrow their shovel. As an engaged field researcher, social capital can be built within the targeted context and with the greater research community as the credibility and validity of research insight are directly connected to social capital~\citep{robert_social_2008}. This social capital is the sum of the researchers' experiences, network ties, technical skills, and resources~\citep{bozeman_scientists_2004}. Accruing social capital emphasizes bringing together people with similar backgrounds and focal concerns. Over time, a researcher can experience privileged access that enables a deep understanding of how a community or organization functions.  

\subsection{The Role of Common Infrastructure in Obscuring Boundaries}

Common technical and practice infrastructure \citep{star_ethnography_1999, star_steps_1996} can make boundaries within open source software challenging to see. From a distance, most people with a passing interest can observe that nearly all open source software today shares a common technical infrastructure that includes Git-based version control tools for managing distributed contributions to software repositories, issue trackers, and varying levels of documentation \citep{mcdonald_performance_2013}. Most people, however, even within open source, cannot state where, precisely, the boundaries between any two sub-networks of open source software exist. The boundaries are perhaps more observable through structured open source software institutions like the Linux Foundation, which generally hosts and supports open source projects in the corporate context. Yet, in other contexts, like open source scientific software, the boundaries can be slightly vague and include some ambivalence toward boundary creation. Yet, routine work across the common technical infrastructure of open source software leads to questions about where one sub-network begins and another ends. Tools like Git, Git repository platforms like GitHub or GitLab, issue trackers, and mailing lists obscure distinctions. For example, scikit-learn is a Python library for machine learning that leverages many other libraries, like matplotlib and NumPy \footnote{https://scikit-learn.org/stable/}. It is tempting to characterize scikit-learn's overlap as a trait that makes it a "boundary object" \citep{star_institutional_1989} or an object that exists on the boundary between sub-networks. Lee \citep{lee_boundary_2007}, and others have suggested this is an oversimplification in some cases and that boundaries between sub-networks can be vague and subject to constant negotiation through what she described as \emph{boundary negotiating artifacts}. 

The well-established common technical and practice infrastructure across open source software is highly ordered and broadly distributed. It has all the critical marks of routine work: self-explanation, inclusion, compilation, structuring, and borrowing \citep{star_institutional_1989}. Nonetheless, open source software researchers question, with some regularity, the boundaries of the sub-networks they choose to define and explain. Therefore, we will stop short of characterizing our questions as \emph{boundary identification artifacts} and instead focus our attention on how social capital built by people in one open source community, as applied as significant for bridging engaged field research into a new and distinct open source sub-network. 

\subsection{Bridging and Bonding Social Capital}

The construct of social capital holds two forms in prior literature: bonding and bridging \cite{pyuqing_ren_applying_2007}. Bonding social capital accrues based on the researcher's engagement in a domain, and the bond deepens as a researcher and a community get to know one another over time. Bridging social capital is developed through a need within a bonded community for outside relationships, expertise, or perspective. At first, there is less clarity about how the new community can apply a researcher's knowledge. Bridging social capital from one community to another can provide the foundation for developing bonding social capital in that new community~\citep{robert_individual_2009}. As such, bridging social capital means that without first-hand knowledge, social capital from similar contexts can provide sufficient social capital for introduction into a new community. However, it will be replaced over time by observational knowledge and new bonding social capital~\citep{robert_individual_2009}.

Social capital is not fungible like currency \citep{coleman_social_1988} but is a critical dimension of engaged field research. Researchers earn, develop, and leverage social capital during long-term, committed research engagements. Effectively balancing the accumulation and leveraging of social capital combines the bridging and bonding types of social capital in different ways over time. Commitment to and the development of bonding social capital in one community comes first, and opportunities for the outcomes of that research to evolve into a bridging form of social capital come second. Bonding social capital provides access to rich data and insights within one community and enables bridging social capital within adjacent communities. To the extent our methodological approaches enable adjacent bridging across contexts, we are more likely to avoid the disconnection of scholarly insight from practice. 

\section{Methods: Bridging from corporate Open Source to Scientific Open Source}

Our methodological approaches use well-recognized, engaged field research and social computing methods. Getting epistemologically valid field access is a piece of hidden work in social computing scholarship, which we seek to make visible. In the course of doing that, our research uses a type of engaged field research specified through soft systems methodology \citep{checkland2006learning, reynolds_soft_2010} and trace ethnography \citep{geiger_trace_2011, goggins_context_2013, goggins_creating_2013}. While conducting our research, we are attentive to transparency with our informants \citep{etherington_ethical_2007} and the complexity and ambiguity of the OSS context \citep{clarke_situational_2003}. We deploy strategies for ensuring that the perspective presented in our research is more resonant than dichotomous \citep{haraway_situated_1988}, and reports from the field not as though it's an outside place \citep{taylor_out_2011}, but instead ensuring we are reporting as authentic a view in the research we produce \citep{haraway_situated_1988, barad_meeting_2007}. 

\subsection{Our Ethical Guideposts for Open Source Software Field Engagement and Social Computing}
Our work on corporate open source is guided ethically by our values as researchers and people to be our best selves. Our work in this space has led us to contribute to numerous corporate open source projects, including co-founding and managing an open source project hosted by the Linux Foundation. Through our deep open source engagement, we recognize and strive to create humane documents and technologies \cite[67-68]{nardi_information_2000}, perhaps most clearly visible in documents and technologies that directly impact individuals. Fritz Lang's observation about the moral of his film, "Metropolis," that the mediator between the brain and the hand must be the heart, as underscored by \citet[67]{nardi_information_2000} is at the center of the information ecology we are building with our collaborators in OSS. Our respect for the potential vulnerabilities of individuals we work closely with has, to date, prohibited the use of real names and personal identifiers in our research publications. However, many individuals identify themselves as contributors to the open source project that we manage \citep[256]{bruckman_when_2015}.

\subsection{(Re)-Entering the Field}
We believe that research in complex sociotechnical systems contains "different stakeholders, with different world views, acting purposefully" in situations where demarcating boundaries is difficult \citep[p. 3-6]{checkland2006learning}. This belief, as derived from Soft Systems Methodology \citep{checkland2006learning}, conceptualizes a lens for understanding engaged field research as positioned within collaborative networks, comprised of various stakeholders that drive actions leading to more effective practices from their point of view. As such, we approach our work with scientific OSS as an act of re-entering the field. The results of our work in corporate OSS described below are a connection point to scientific OSS organizations. 

Our history of accruing bonded social capital necessary to be effective in the corporate open source context is deliberate. We recognized that institutions like the Linux Foundation, a registered trade organization, help coordinate between open source projects and the thousands of member technology firms under their umbrella. For example, the Linux Foundation originated to ensure the Linux Kernel would be consistent across distributions because market differentiating software runs on Linux, and managing different product builds for different Linux kernels would be more costly. Our work with the Linux Foundation began in its early stages, and our recognized bonded social capital accrued based on our prior contributions to open source software projects \citep{germonprez_risk_2009,germonprez_organizational_2011, graves_sifting_2014, mcdonald_performance_2013,mcdonald_modeling_2014}

Beginning with an NSF VOSS grant, author Tango progressively accrued their bonded social capital within the Linux Foundation toward more of the insider views that emerged through project-specific contributions and a series of studies that produced both scholarship and valuable insight for practice  \citep{germonprez_dosocs_2016, kendall_game_2016}. Work with the SPDX and FOSSology projects was beneficial for building a common language between author Tango and the respective communities. Further, author Tango leveraged early open source engaged field research into the co-founding of a Linux Foundation brokered open source project to provide meaningful insights regarding the health of open source projects. Likewise, author Foxtrot leveraged their NSF VOSS grant and funding from the Navy and Department of Education to build bonded social capital in open source more broadly construed, and ultimately, the Linux Foundation. Author Foxtrot's multi-disciplinary and systematic integration of qualitative methods with trace data analysis accrued social capital with Linux Foundation member corporations through their ability to quickly develop and deploy metrics regarding open source community health in a corporate context. Author Tango's and Foxtrot's work in corporate open source resulted in an accrual of bonded social capital, contributing to an ability to use standard open source technologies and philosophies to provide swift, helpful information and provides bridging social capital as we move into other networks (i.e., scientific open source software).

\subsection{Trace Ethnography}
As we seek to re-enter the field, a second method of inquiry we used is trace ethnography \citep{geiger_trace_2011}, which aims to build context around the OSS trace data we gather by analyzing Git logs and Git Platform API's like those on GitHub and GitLab. These traces of individual interactions with and contributions to collections of repositories identified by informants bound each collection as a distinct OSS sub-network. Trace data are triangulated with previously identified patterns in OSS project structure and tuned to align the estimation of connection between a contributor and a project with how our informants describe their engagement with projects during our prior fieldwork \citep{goggins_group_2013, goggins_connecting_2014}. Trace ethnography provides a succinct, empirical validation that corporate and scientific open source software are manifest as distinct sub-networks. To identify the network structure of each, we operationalized commit and platform message metrics as weighted connections between individuals and projects in two scientific OSS sub-networks of ..... and one corporate OSS sub-network of ..... Our use of trace ethnography ensured that our movement between unique sub-networks was, in fact, exceptional. 

Our trace ethnography reflects our prior field data and a connection weighting method \citep{goggins_group_2013} that directly incorporates insights gleaned from our previous research. For example, our informants noted that code contributions reflect a more robust connection than platform messages around issues and merge requests. We wove this insight with trace data to value connections derived from code contributions more heavily than connections derived from messages. We also used the time distance between messages to decrease connection weight as the time distance becomes more significant. In addition, the message content is evaluated in our weighting, with topically and linguistically similar text in sequential messages being given more weight than less congruent messages \citep{dodsworth_language_2019}. Finally, we account for longitudinal patterns \citep{duxbury_longitudinal_2022} by weighting actions from the most recent 18 months of activity more heavily than older interactions. The result is an expression of social network centrality measures \citep{carrington_models_2005} that produce visually succinct structural models of the communities identified within each ecosystem using a combination of Leiden \citep{traag_louvain_2019} and Louvain computational models \citep{murniyati_expanding_2023}.   

\subsubsection{Structural Contrast of Corporate and Scientific OSS Sub-networks}
\begin{figure}[hbt!]
    \centering
    \begin{subfigure}[b]{0.3\linewidth}
        \includegraphics[width=\linewidth]{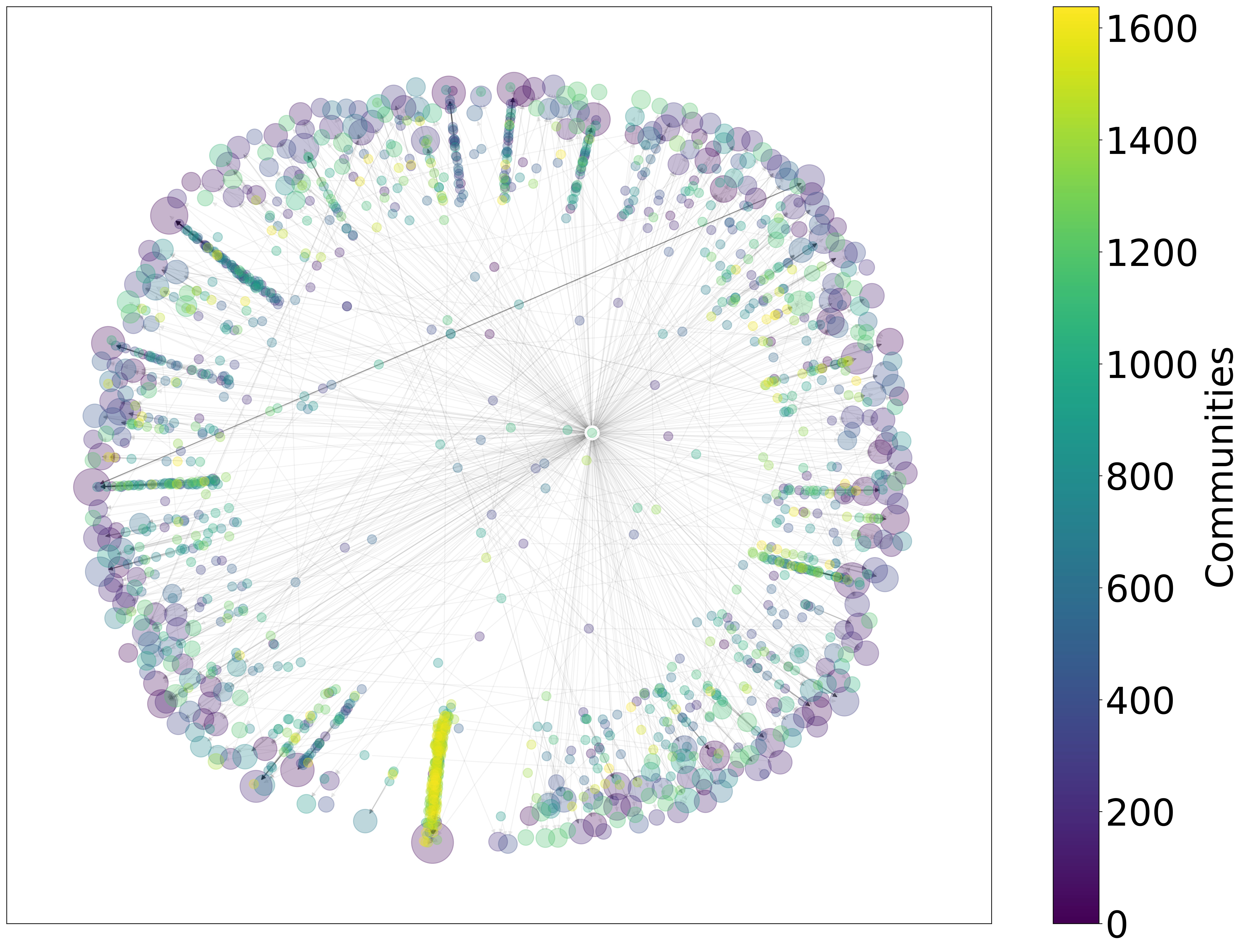}
        \caption{Corporate: Donut-shaped structure.}
        \label{fig:x corpnet}
    \end{subfigure}
    \begin{subfigure}[b]{0.3\linewidth}
        \includegraphics[width=\linewidth]{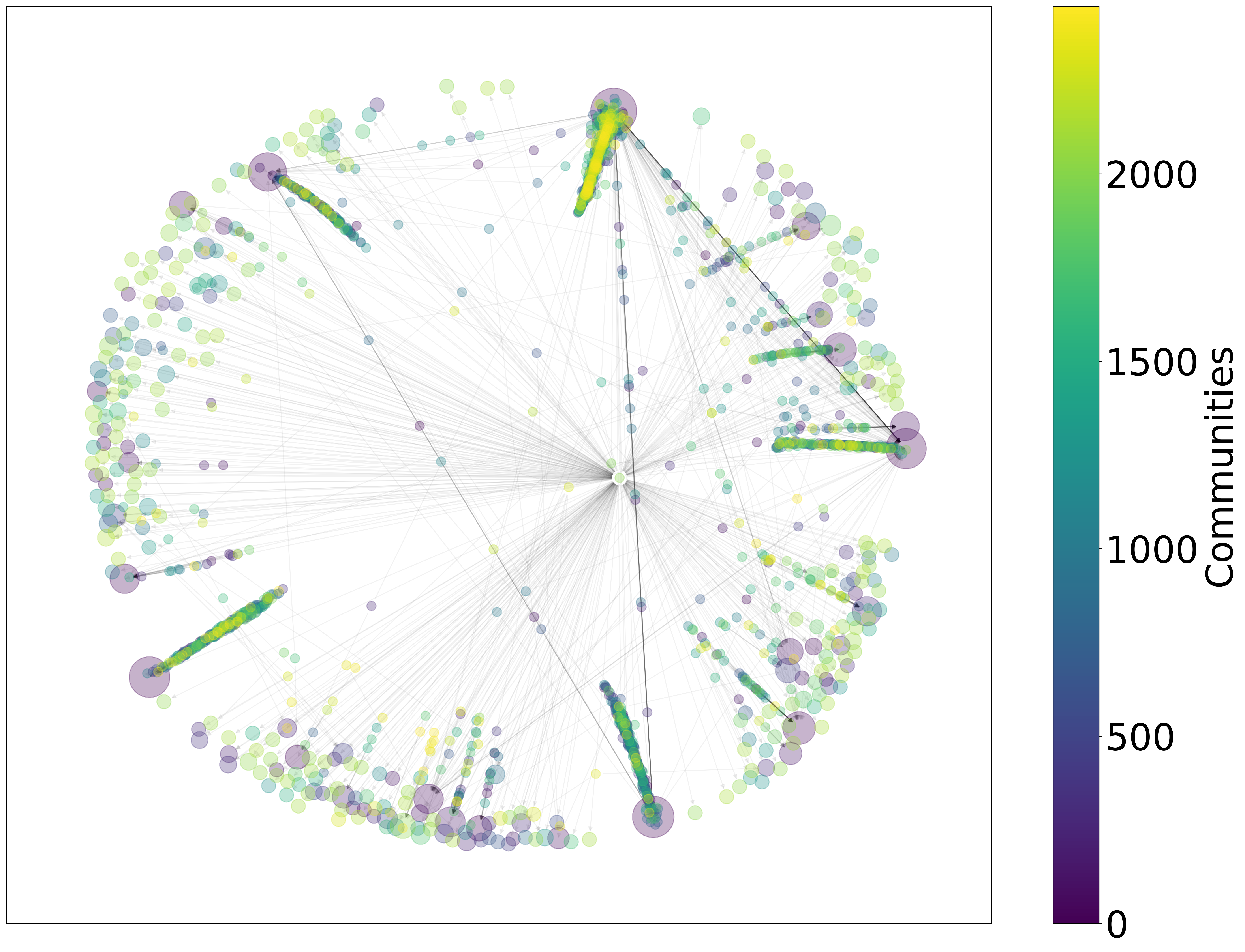}
        \caption{Funded scientific: Clock-shaped structure.} 
        \label{fig:x czi}
    \end{subfigure}
    \begin{subfigure}[b]{0.3\linewidth}
        \includegraphics[width=\linewidth]{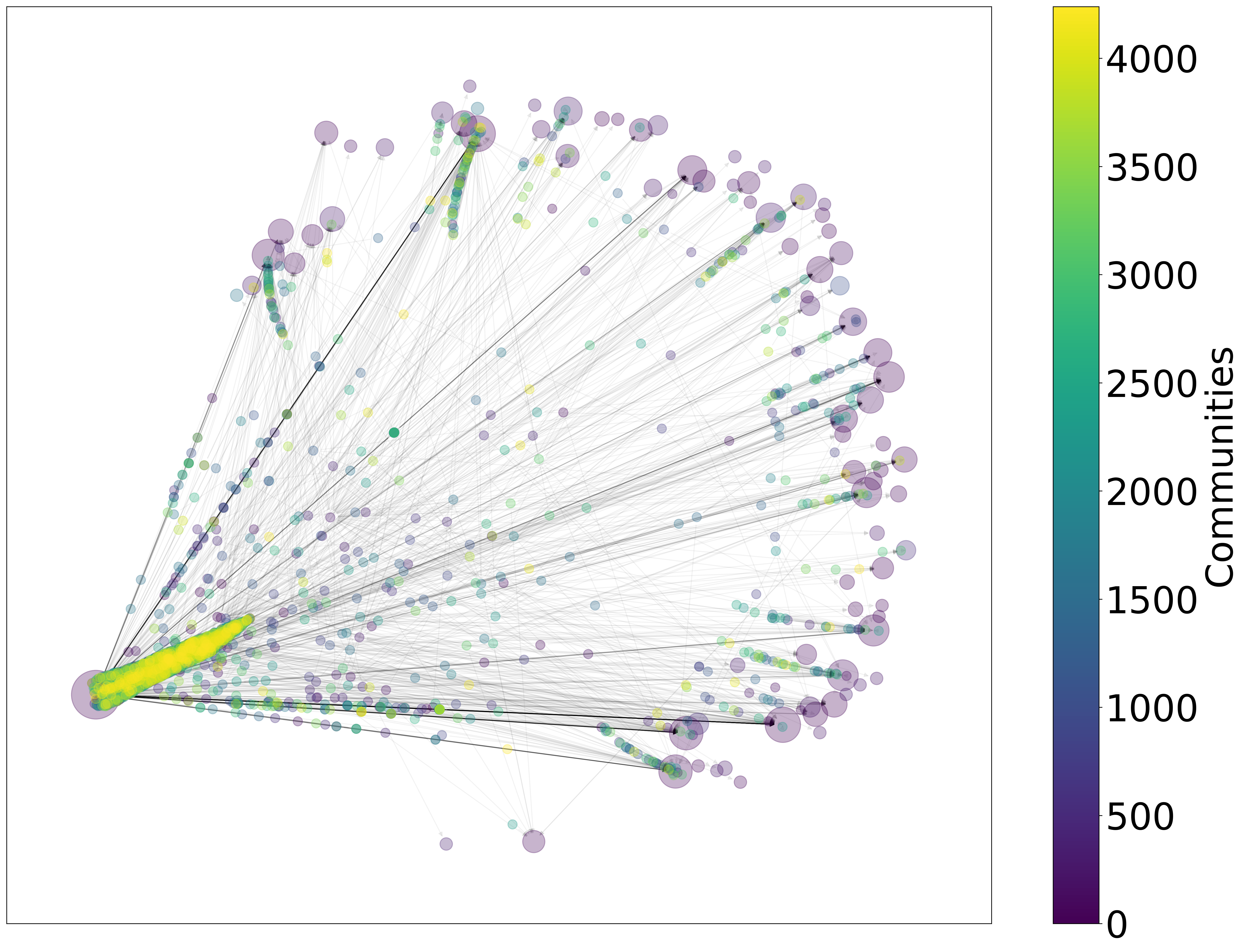}
        \caption{Organic Scientific: Funnel-shaped structure.}
        \label{fig:x harvardsub}
    \end{subfigure}
    \caption{These are sociograms of the most prominent communities within a
        corporate (\subref{fig:x corpnet}), 
        funded scientific (\subref{fig:x czi}), and
        naturally occurring scientific (\subref{fig:x harvardsub}) open source software contribution networks. These differences in structure require different approaches for engaged field research, while the common central artifact of a software repository and motivation to understand the relative health and sustainability of individual repositories within each ecosystem frame the methodological importance and points of transferability of social capital from one context to the next. 
    } 
    \label{fig:threegraphs}
\end{figure}

Reflexive, qualitative analysis of metrics, weighted using insight from field notes, illustrates structural differences between the sub-networks. This method surfaced different patterns of structural organization, which, combined with our field notes, made it clear that there are, in fact, material differences in how each sub-networks contributors engage with different repositories. This process is productive for us, as researchers, and engaging to our field sites because they, too, do not want to have statistics without underlying context. Figure \ref{fig:threegraphs} illustrates a doughnut-shaped structure for the corporate OSS sub-network, a clock-shaped structure for a funded scientific OSS sub-network, and a funnel-shaped structure for an organic scientific OSS sub-network. Node colors identify individuals in distinct communities, with all repositories in "community 1" and purple. The connecting lines and sometimes linear alignment of contributors around a particular community indicate levels of dominance of one collection of projects over others in a sub-network. Dozens of these communities are visible in the doughnut-shaped corporate OSS sub-network.

In contrast, the clock-shaped funded scientific OSS sub-network has 5-6 major community clusters, and the funnel-shaped organic scientific OSS sub-network is visibly dominated by a singular community around a single, dominant project (in this case, Bioconductor). Larger versions of these sociograms can be examined in section \ref{sec:example-section}, Appendix A. Other specific visual characteristics to note include: 
\begin{enumerate}
	\item  In-degree centrality is reflected in node size. Since only the repositories have In-degree centrality above 1 in our data organization, they are the only nodes that appear larger. 
	\item The distance between nodes in each graph is determined by the similarity of contributor patterns toward each of the repositories from the three sub-networks.
	\item The shapes of each sociogram reflect the structure of each expert-identified sub-network. 
\end{enumerate}
These differences illustrate different practices and interconnections that define these sub-networks and succinctly illustrate that the social and contribution structures of scientific and corporate OSS sub-networks are empirically different. The individual and collective motivations and rewards of engagement in each sub-network are distinct. As such, these differences in structure require different approaches and uses of social capital for engaged field research between sub-networks.

\section{Findings: Social Capital in Spanning Sub-networks}

Our approach to working across sub-networks relies on our experience in helping open source communities and organizations understand the OSS health indicators (i.e., success factors, risk, and sustainability) of projects they rely on. Failures within upstream open source software networks can have negative consequences for people and organizations relying on their output. Developing metrics and evaluating the sustainability of OSS projects in a corporate context enabled us to communicate similar concerns later in a scientific OSS sub-network. However, entering the domain of scientific OSS without this experience would, ultimately, not be very different from entering a new field site. As such, we were centered on a long game of understanding and usefully contributing to open source software in different sub-networks, enabling a deeper understanding of OSS health in all of them through the revelation and explanation of shared and distinct practices. In each sub-network, we focused on learning the language, contributing, and assuming leadership through a series of systematic steps. In this section, we highlight the approach by which we successfully moved from one sub-network to another, describing our engagement in corporate OSS and how we bridged it to scientific OSS. Finally, we reflect on the role of technical utility in our experiences and present a synthesized summary of how technical utility acts as a bridge between sub-networks that exist in technically parallel and sociotechnically orthogonal worlds to re-conceptualize the social and the technical in both. 

\subsection{Engaged Field Research to Establish Bonding Social Capital in Corporate Open Source}

Our understanding of corporate OSS accrued a deep knowledge of the social engagement and technical practices within that sub-network. Through this work, our ability to provide insight and value to the communities we engaged with grew over time. This corporate engagement culminated in the creation of a new Linux Foundation open source project - the  Community Health Analytics Open Source Software (CHAOSS) project. The CHAOSS project originated from an "unconference" session at the Open Source Leadership Summit 2017. The session, funded through a Tango-Foxtrot's collaborative grant from the Alfred P. Sloan Foundation, started our deep investigation of OSS health. 

The session revealed how corporations actively view and evaluate engagements with open source projects and how the health of these projects and the communities of people who build and maintain them were at the top of their minds. Following the session, CHAOSS was established at the Linux Foundation to combine efforts to understand the health of open source communities and create software tools to extract and visualize sustainability data. Tango and Foxtrot were instrumental in creating CHAOSS and remain active contributors to the project as board members, maintainers, and contributors. 

Since its inception, the CHAOSS work of Tango and Foxtrot has had a strong presence at open source conferences: Open Source Summit North America (2017, 2018, 2019, 2021, 2022, 2023), Open Source Summit Europe (2017, 2018, 2019, 2022, 2023), CHAOSScon Europe at FOSDEM (2018, 2019, 2020, 2022, 2023, 2024), CHAOSScon North America (2018, 2019, 2020, 2021, 2022, 2023, 2024), Open Source Leadership Summit (2018, 2019, 2021, 2022, 2023), and the Community Summit (2018, 2019, 2021, 2022, 2023). Attendance at CHAOSS sessions at these conferences and contributions to the project has continually grown. Interest has come from corporations, open source foundations, project maintainers, and open source project members, indicating that this topic is essential to many. This work applies traditional research methods to construct academic research outcomes \citep{germonprez_eight_2018, germonprez_rising_2019}, yet considerable time was also spent engaged in community building and community contributions. The CHAOSS project's principle aims to make open source project sustainability more transparent and actionable for community members, providing direct value to the open source community members interested in understanding project sustainability. 

The CHAOSS metrics developed for measuring open source community sustainability generated indicators describing how groups around any open source project function. For example, the CHAOSS project's published metrics include indicators of how many different individuals contribute to the project over time through standard open source technical components as software commits, merge requests, issues, documentation, and communication in the forms of comments on each of those artifacts, as well as participation in mailing lists. We developed the Augur software to examine open source software projects, and we can quickly generate CHAOSS metrics for any software project or collection of projects. In this way, the infrastructure of the CHAOSS project affords us the technical foundation and bridging social capital outside the corporate open source sub-network, where our work originates. 

Within the CHAOSS project, we deeply understand the social arrangements and technical practices manifest across corporate OSS as members, contributors, and researchers. Outside the CHAOSS project, our capacity to understand adjacent open source contexts, derived from understanding elemental tasks like issue creation and code creation, provides a bridge. While we have extensive sociotechnical expertise in our corporate open source sub-network, our ability to perfectly translate that expertise to a new domain is inhibited, not because of our technical skills but because of how identical OSS technologies are operationalized and socialized in scientific software practice. Our comprehension of social and technical phenomena is not zero in these new contexts (horizontally) because they use identical OSS technologies. Therefore, our move to new contexts requires a commitment to (1) effectively building bonded social capital by demonstrating technical utility, and then (2) adapting our sustainability expertise to account for the distinct patterns of engagement found in open source scientific software.

As directly crucial to our deepening understanding of social engagement and technical practices, in our time at the CHAOSS project, we found that making OSS health data transparent and actionable was applicable in the corporate open source sub-network and other OSS sub-networks. That is to say, open source community health is relevant to many different sub-networks, and the tools and methods deployed to understand community health may provide insight across many different contexts, as illustrated in Figure 2 ~\citep{germonprez_rising_2019, germonprez_eight_2018} CHAOSS provides utility to OSS stakeholders by defining a taxonomy of community health metrics. That same taxonomy, consistent if not perfect, and the CHAOSS software, \citep{goggins_augur:_2021}, implements concrete analysis using CHAOSS metrics served as bridging social capital to cross into new OSS sub-networks, with scientific OSS being a most prominent example.   

\begin{figure}
  \centering
  \includegraphics[width = 100mm]{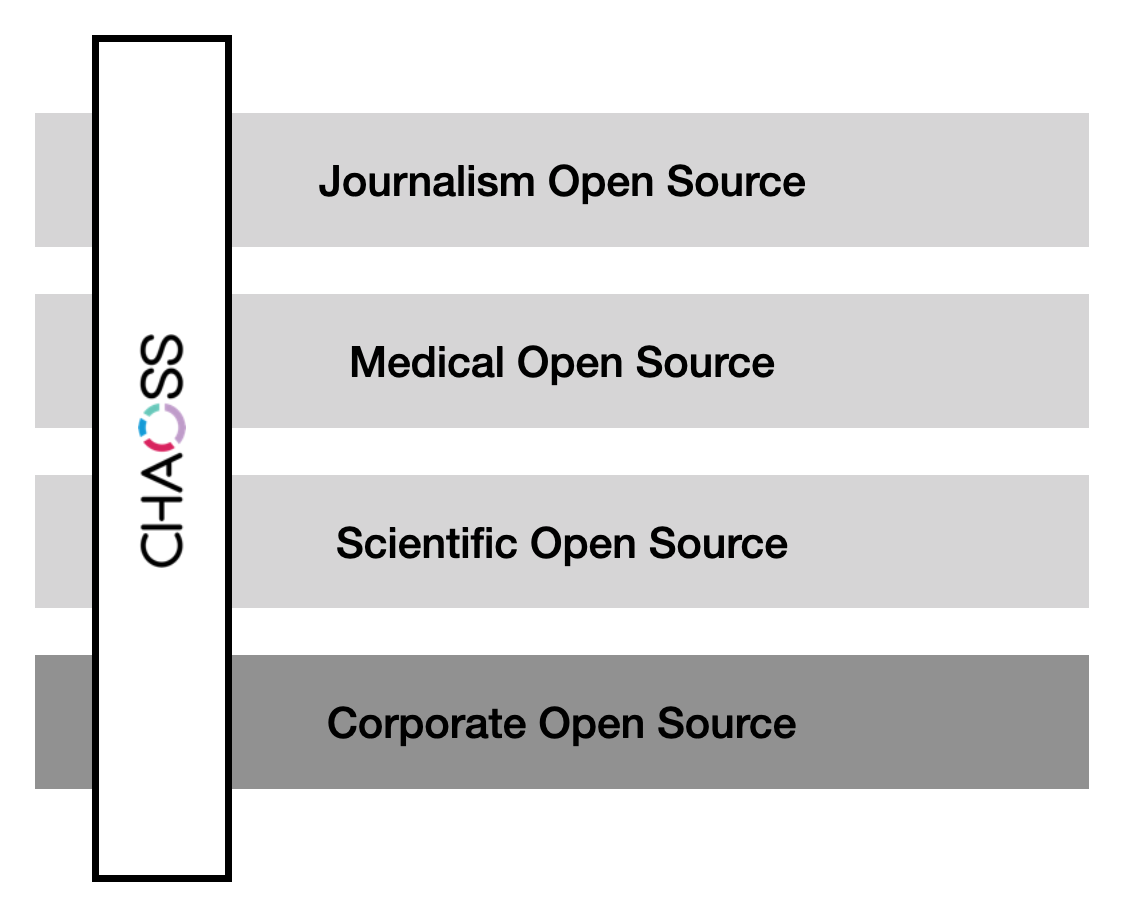}
  \caption{}
  \label{fig:chaoss}
\end{figure}

\subsection{Engaged Field Research to Apply Bridging Social Capital to Scientific Open Source}

To some extent, our research focus on open source across sub-networks relies on the barely fungible currency of social capital \citep{coleman_social_1988}. Without first-hand social knowledge, bonding social capital from other contexts can introduce social capital in new areas \citep{yuqing_ren_applying_2007}. This initial social capital is bridging social capital. However, over time, bridging social capital must be replaced by bonding social capital as new observational knowledge is made available to community members~\citep{robert_individual_2009}. Scientific open source exists at the intersection of four focal context areas - software development, open source communities, scientific research, and domain-specific areas (e.g., Biology). In working with scientific open source, our research team brought bonding social capital from other contexts (e.g., corporate open source), providing us a point of entry. However, we knew we needed to understand this new sub-network and develop the language to understand and ultimately contribute to the goals and knowledge of new communities of interest (e.g., scientific open source and domain-specific areas). Our bridging social capital via CHAOSS was fleeting~\citep{robert_individual_2009}, and it needed to be replaced in this new context by bonding social capital to maintain and grow and establish ourselves in a new sub-network. 

It was only at this point where we began to see the role of both bonding social capital and bridging social capital~\citep{coffe_toward_2007, patulny_exploring_2007,goggins_group_2013, yuqing_ren_applying_2007} to construct the relationships necessary for effective fieldwork in new sub-networks. Our social capital, developed through bonds in corporate open source software, granted us some degree of bridging social capital into open source scientific software because of a shared goal of understanding open source community health. Our bridging social capital was derived, in part, from the dimensions of our long-term field engagement with OSS and emerged from an opportunity to examine open source scientific software projects and their expressed needs, vision, and heuristic definitions of OSS health. 

\subsection{Five Stages Of Building Social Capital in the Scientific Software Sub-network}

First, researchers must be committed to a orienting themselves to understand adjacent sub-networks~\citep{rich_coal_2013}. It is reasonable for researchers to engage in programs that do not require much investment in accruing this domain knowledge and still produce high-quality research. Knowledge of new sub-networks demands intentionality motivated by a researcher's long-term goals. While each personal research program informs the size of this investment, some significant investment is necessary because building knowledge and credibility in an adjacent sub-network depends on two parties knowing each other well over time~\citep{lacity_review_2009}. Open source engagement can mean joining an open source project as a newcomer - downloading source code documentation, following the community discussion on mailing lists and issues forums, and posting questions and comments relevant to the community~\citep{jensen_joining_2011}. As relational ties strengthen, bonding social capital is accrued through a commitment to the norms, values, and beliefs of a community~\citep{chua_enacting_2012}.   

Second, a researcher must accurately map the new sub-network to accrue an understanding of the specific idiosyncrasies necessary for engaging with nuances found in any new setting. Context mapping is not an exercise returned to the context in question. Instead, it is an activity for a researcher alone. Dimensions of mapping can include a review of relevant literature germane to the context,  familiarity with the actors and activities present in the context, and computational mapping representing how projects within a sub-network or tied together. Bonding social capital can accrued with understanding and sharing these findings with a sub-network~\citep{lacity_review_2009,coffe_toward_2007}. While a researcher will advance research and a community's goals that they engage with, the transaction cost of mapping a context is high. It is a knowledge-intensive activity requiring high domain knowledge, experience, and intensive learning~\citep{von_krogh_community_2003}. These costs hinder many researchers because context mapping does not immediately result in key research or practice contributions.  

Third, a researcher must build language skills to understand and contribute to the new sub-network. Communities often develop specialized terminologies or jargon that signal community or "in-group" \cite{turner_social_1979} membership. Language skills signal "in-group" membership, developing over time and ensuring the researcher's familiarity with key concepts and phrases in the study context ~\cite{agerfalk_outsourcing_2008} that strengthens engagement in a new sub-network. Over time, like mapping the new context, developing language skills is a transaction cost for the researcher. However, it leads to greater knowledge of the new sub-network, which, in turn, enables increased access to more significant insights, which is likely impossible for researchers who do not understand the specialized language.  

Fourth, a researcher must contribute tangible value that community members recognize. Deeper insights, derived from such contributions to a group~\citep{turner_social_1979} at this stage, enable detailed narrative description, case studies, empirical analysis, and theory development beyond mere tertiary or secondary means of engagement~\citep{coffe_toward_2007}. Providing value to the community builds a researcher's credibility and utility~\citep{lacity_review_2009}, providing researchers with access to rich data~\citep{robert_social_2008}. By contributing to the study context, researchers acquire information that can lead to more significant insights than previously published. This accumulates access, insight, and previously unattainable knowledge within a sub-network. 

Fifth, a researcher can then again leverage their utility to an engaged research site beyond their immediate context to bridge other adjacent areas of interest. For example, scientific OSS is sometimes called research OSS, a broader umbrella that includes both arts and science. Moving beyond the first adjacent research context helps to build relationships in communities of varying heterogeneity~\citep{coffe_toward_2007}. Understanding new research contexts in adjacent domains is derived from long-term contribution and acknowledgment in one community where the engaged field researcher has contributed, allowing them to bridge to new contexts. The materiality of what generates such connection may be deep knowledge of an adjacent context \citep{nardi_context_1996, blincoe_leveraging_2012, goggins_context_2013}, capacity to measure outputs from a common infrastructure \citep{blincoe_proximity_2012}, or individually formed trust bonds between researchers and practitioners working in mutually-interested fields \citep{valetto_actionable_2012, goggins_creating_2013, goggins_assessing_2014}.

In some ways, it may seem we are arguing that commitment, context assessment, language skills, contributions, and bridging to new contexts are five stages through which one might begin to doubt Coleman's~\citep{coleman_social_1988} assertion that social capital is not fungible. Our argument, instead, is that treating knowledge of adjacent sub-networks as fungible neglects to recognize the invisible work of building knowledge of new networks. Can knowledge of adjacent networks be accrued and expended in transactional ways? We think the literature suggests the answer to both questions is "no ." Although adjacent contexts often share a common technical infrastructure, knowledge of adjacent networks may appear more transactional. However, credibility in an adjacent network is inherently necessary for negotiating third-order infrastructural challenges~\citep{star_steps_1996}. The deep insights sought by long-term engaged field researchers and the accrued knowledge of adjacent networks aim to theoretically and practically reveal the complex tensions in sociotechnical arrangements.

\subsection{Case Example: Establishing Ourselves in the Scientific Open Source Sub-network}
\subsubsection{Case Example: Orientation}

Our entry into the scientific software sub-network began with the examination of detailed road maps of nearly 300 scientific open source projects. At this point, our research methods centered on learning the landscape and language through a detailed qualitative content analysis of funding proposal documents submitted by open source scientific software projects. In these documents, open source project leaders identified their challenges, prior successes, and individual goals. Insight from our prior experiences in open source software development allowed us to synthesize these concepts into recurring themes in open source scientific software development. The documents' open and axial coding identified concerns associated with developing scientific open source software. We constructed several themes relevant to the health of open source scientific software. Specifically, the following issues emerged consistently around the theme of contributor constraints in open source scientific software: 

\begin{enumerate}

\item Having sufficient contributors available to maintain and improve the software. 
\item Ensuring a large enough set of core contributors for the project to onboard new contributors promptly.
\item Overcoming challenges associated with poor documentation impeded new contributor on-boarding. 
\item Building a large enough community of contributors to recruit and train new contributors in more corporate environments. 

\end{enumerate}

Our findings at this early stage indicated that scientific open source projects are concerned about a lack of contributors and have difficulty building communities of contributors. Compared with more traditional volunteer and corporately engaged open source projects, there are fewer contributors to the projects available from the open source scientific software projects we examined. The low contributor counts likely drive concerns about growing demands from software users for new features, complete software testing, and research lab-specific data integration support. Many scientific open source software projects expressed risks like collapsing under their weight, resulting from growing user demands and a shortage of contributors and time. Interestingly, while scarcity of contributors was an identified and known concern, most of the community outreach and recruiting activity proposed by these projects was focused on building user bases rather than contributor bases. 

To triangulate our qualitative observations, we empirically validated concerns about contributor levels. Our question was: Are there \emph{really} fewer contributors on open source scientific software projects than more corporate projects? To do this, we compared the 2,035 publicly available open source repositories from nearly 300 scientific open source projects we reviewed against 6,744 corporate OSS projects sampled from companies that work in open source. We looked at several factors and hypothesized the challenges articulated by open source scientific software project leaders' differences in participation diversity and size of the contributor base for those projects compared with corporate open source software. Specifically, open source scientific software repositories have nine or fewer committers in 60\% of the 2,035 repositories examined.

This effort enabled us to identify similarities and differences between corporate and scientific open source sub-networks, helping us establish what we know and don't know and how to best move forward. We knew that our knowledge of corporate OSS could not, by fiat or process, automatically transform into a well-fitted utility for this new scientific context. To move forward, we used five phases that we developed in our corporate open source efforts as guides to help in our entry into this new sub-network. As existing field engagement methodologies do not speak explicitly about the role of social capital in field research, we suggest that these five phases lead one to a conceptual model of the steps for moving between sub-networks.

Additional orientation efforts included regular participation in two scientific software organizations following our initial deep engagement with a collection of scientific projects we described earlier. First, we participated with "FAIR for Research Software" \footnote{\url{https://www.rd-alliance.org/groups/fair-4-research-software-fair4rs-wg}}, whose aims are focused on helping scientists who happen to build open source software develop a shared path for recognition of those contributions within their institutions and the scientific community. In addition, FAIR for Research Software also helps scientists who develop open source software learn about five critical characteristics for their projects to strive for in the interest of sustainability: 1) have a version-controlled repository, 2) include an open source license, 3) register the software in a community registry like CRAN, for R-based statistical software, for example, 4) enable citation of scientific software, using publications like the Journal of Open Source Software \footnote{\url{https://joss.theoj.org/}} or using markers like those available through CiteAs \footnote{\url{https://citeas.org}}, and 5) use a software quality checklist \footnote{\url{https://fair-software.nl/recommendations/checklist}}. Second, we participated in the Research Software Alliance (ReSA) \footnote{\url{https://www.researchsoft.org/}}, whose mission aligns with FAIR4RS but aims more directly at the general advancement of research software quality in OSS. Our collaboration focuses on developing a large, in-person workshop in mid-2022 and seeking specific mechanisms for applying, sharing, and informing members about using CHAOSS metrics to advance scientific OSS. 

\subsubsection{Case Example: Mapping Open Source Scientific Software}
We built our map of the open source scientific software landscape from two perspectives to focus on accruing knowledge of the open source scientific software sub-network. First, we expanded our analysis of the aforementioned set of 2,035 scientific OSS repositories using an integration of six different computational modeling techniques: Clustering, topic modeling, computational linguistic analysis (CLA) of discussion novelty, CLA of sentiment, predictive modeling of pull request acceptance likelihood using our specialized combination of statistical models, and a random forest algorithm to identify anomalies in the statistical rates of participation for each repository in our previously noted corpora. We combined those results into a grouping of 14 collections in the original corpora and are using the results as we continue our field engagement with FAIR4RS and ReSA. Second, we analyzed 92 scientific OSS repositories from the US National Science Foundation-funded projects. In this work, we are collaborating with CiteAs to mutually understand the characteristics of funded projects that resulted in sustained scientific OSS projects and those that did not, using CHAOSS metrics. We also (1) seeded our list with projects we are familiar with through our initial engagements and prior research and (2) generated CHAOSS metrics for larger sets of scientific OSS projects, generating a high-level understanding of the health of those projects and how the open source scientific software space is structured. Our survey activity emerged as an exemplar of how knowledge of adjacent sub-networks develops through a combination of the bridging social capital derived from expertise and usefulness in an adjacent domain. 

\subsubsection{Case Example: Building Language Skills}

From our outreach and mapping efforts, we focused on contributing value back to the open source scientific-software sub-network. To do this, we leaned on our understanding of the language in the corporate open source sub-network so we could effectively learn and contribute value back to people in the scientific sub-network. We built our language in this new sub-network by attending meetings, conferences, and workshops aimed at scientific open source software. We met with people directly in their communities to learn about their challenges with respect to the development of their scientific software and the management of their communities. In our focus on building language skills, we found that much of our ability to bridge between corporate open source and scientific open source stemmed from how fundamental open source concepts are used. This includes, for example, language around community responsiveness, licensing and copyright, newcomer experiences, and contributor retention. What was unique when bridging between sub-networks is how concepts are drawn together in different ways to achieve meaningful goals for a respective sub-network. Knowing how to assemble concepts in meaningful ways for the open source scientific-software sub-network turned out to be the primary focus of our language building, developing an awareness of how members attend to open source concepts as applicable to goals around community growth and decline, stewardship and sustainability of communities, software maintenance, and ecosystem management. While we could rely on fundamental open source concerns that we learned in our engagement within the open source corporate sub-network, we were required to reassemble our use of the concepts in meaningful ways in the open source scientific-software sub-network. 

\subsubsection{Case Example: Contributing Value}
Combining our engagement with FAIR4RS and ReSA we identified a subset of exemplar projects in each organization. Our work here focused on the impact and scope of those projects within our larger corpora of several thousand scientific open source software projects described earlier. We analyzed connections between the projects by identifying contributors who participate in the exemplar projects and other projects inside and outside our larger corpora, with preliminary findings suggesting the existence of programming language-centered "contributor groups" within open source scientific software communities. The exemplar projects helped map the space of scientific software, which is less organizationally bound than we find in corporate OSS. We hypothesize this is because scientific OSS is more diffuse and varied. These findings are beginning to emerge as helpful information to project owners, who contrast projects they view as successful with their projects. This previously unavailable awareness is beginning to help ground the efforts of open source scientific software project maintainers to improve their software and expand their contributor communities. 

Working orthogonally with our analysis and contextualization of scientific OSS within the exemplar projects, we discovered relationships between the "contributor groups" identified and scientific papers resulting from funded projects in the 92 project CiteAs corpora. This work included identifying related papers through systematic literature reviews focused on criteria of scientific impact in concert with open source scientific software usage. We aim to help the community of open source scientific software builders tease out the networks and technical mechanisms that elevate the impact of their software on science and when, where, and if that impact is visible in publications. 

From our conversations with scientists who build open source software, it is clear that their use of any particular project can be idiosyncratic to their labs, available equipment, and the team member skills working there in a given year. We are investigating a small number of individual cases where this is apparent, and our fieldwork shows these practices to be at once widespread and labor-intensive to understand. Unlike the relationship between scientific OSS and scientific publications, these assemblages exist in isolation. To fully understand how individual laboratory OSS software pipelines follow patterns or validate an early hypothesis that they follow some pattern, we described how the type of long-term field engagement experienced through our work in CHAOSS will generate further insight. Through these processes, we recognize that the knowledge of adjacent networks in open source software will help build understanding in a broad sense of the health of open source scientific software by: 
\begin{itemize}
    \item improving information resources for stakeholders,
    \item identifying critical yet under-resourced open source scientific software and
    \item identifying local collections of open source scientific software as meaningful for scientific scholarship.
\end{itemize}

Throughout our work, bridging knowledge of corporate OSS to scientific OSS is possible because we provide a contributed resource to the communities, projects, and organizations we work with in the scientific OSS ecosystem. Our preliminary findings point strongly to scientific and corporate OSS maintaining sharply distinct characteristics and practices. While work remains, we believe we are already contributing knowledge to scientific OSS communities by helping identify areas of concern that may require attention to maintain the health of their communities. Further, this information creates value by providing a glimpse of systematic and structured scientific OSS community work. 

\subsubsection{Case Example: Bridging to New Sub-networks}

From our work in the corporate and scientific open source sub-networks, we can now bridge to new contexts regarding organizational engagement with open source, including the university open source sub-network. We are now attending to the emerging context of university open source - focused on supporting university activities tied to open source, including research excellence, research translation, education, and community development. Like the relationship between scientific and corporate open source sub-networks, university open source shares fundamental open source similarities that we use to bridge to this new sub-network. Having done similar bridging work between corporate and scientific open source sub-networks, we can bridge to the university context as we can focus more immediately on learning how university open source assembles fundamental open source concepts in ways relevant to their sub-network. To date, our efforts have focused on developing relationships in understanding questions and challenges university open source members face. This has included meeting with university open source members from academic institutions and connecting with existing efforts in this context. The scale of university open source is considerably smaller than that of corporate or scientific open source. Fewer than 30 universities have dedicated efforts to support open source research excellence, research translation, education, and community development. 

\section{Discussion: Building and Using Social Capital for Long Term Sociotechnical Engagement}

This paper explicates our method and process for researching open source software across sub-networks and describes social capital's role in engaged field research methods. Agre \cite{agre_practical_2004} pointed out how centuries of political theory development missed the essential truth that social skills are a fundamental concern in political science \citep{agre_practical_2004, agre1997reinventing}. We illustrated social capital's essential function and utility for examining large-scale sociotechnical infrastructures, which is critical for developing a more robust social computing theory \citep{hammond1988relationship, giorgi1997theory}.

Star and Ruhleder apply Bateson’s~\citet{bateson_steps_2000} discussion of three levels of issues to the development of infrastructure. First-order issues are easy to identify and can be addressed by adding or redistributing existing resources. Second-order issues are more challenging to locate and can arise from the interaction of two or more first-order issues. In the study of open source software, first- and second-order issues center around the technical aspects of these sociotechnical systems. When bridging from one sub-network of open source software to another, "startup knowledge of adjacent social and practice networks" is derived from a current understanding of common problems within these first two orders of adjacent sociotechnical infrastructures. Our knowledge of the shared technical infrastructure of open source software is a materiality \citep{leonardi_materiality_2012} that is one component of our startup knowledge of adjacent networks. Third-order issues are broader in scope and involve deep conceptual disagreements or political issues. It addresses how different types of open source software enact third-order issues through differentiation among social groups \citep{tajfel_differentiation_1978, tajfel_social_1982-1} that comprise a network's social constructs and practices, making this a form of systematic development across contexts a vital methodological concern for the sociotechnical communities.  

The knowledge of an adjacent social and practice network that will need development requires what Star and Ruhleder refer to as understanding the role of infrastructure as an invisible yet vital aspect of system adoption through an analysis of large-scale collaborative systems \cite{star_ethnography_1999}. In characterizing this communicative context, Star and Ruhleder distinguish three levels or orders of issues that occurred in communication during the infrastructural development of a large-scale collaborative system. We found that research on open source software requires the development of knowledge of adjacent social and practice networks~\cite{coleman_social_1988, wenger_communities_1998} within each specific sociotechnical infrastructure (corporate and scientific) to address third-order issues. 

\subsection{Hierarchies for Engaged Sociotechnical Field Research }

Our experience observing the effects of gaining knowledge within an adjacent social and practice network and recognizing its importance to long-term research engagement contributes to existing field research methods in several ways. Context matters when the integration between social and technical processes is developing~\citep{nardi_context_1996}. To elaborate on the functional effects of acquiring knowledge of adjacent social and practice networks in open source software, we assert a contextual hierarchy to recognize, assess, and adapt to in each specific case of building something like social capital in a new research context.  

Embedded field research includes attention to processes, outputs, and existing units of organization, both formal and informal~\citep{alderfer_studying_1982, peterson_embedded_1998}. Longer-term field engagements have some similarities with ethnographic research~\citep{star_ethnography_1999, wolcott_ethnography_1999, mead_cultural_1958}, as the researcher is in the role of participant-observer in some moments. However, unlike ethnographic studies, the researcher is more frequently engaged as a facilitator or participant in transformative actions within the research context~\citep{stringer_action_2013}.  

To reiterate a key idea from our work, from a distant perspective, open source software might appear to be a singular context, with nominal domain and contextual differences. However, our work exposes social distinctions across these contexts with implications for the researcher's process to ensure effective engagement, leading to new knowledge and improved circumstances for each examined network. For example, within the corporate open source network, our long-term engagement and research program now rely heavily on our deep understanding of the domain, which results in shared success for us as scholars and informants. One could think of that sort of knowledge of adjacent social and practice networks as “top-down” because, like ethnographers, we fully embed ourselves with support from organizational leadership, and we are building a long-term record of effective organizational change.  

As we learned in the open source scientific software community, bridging our knowledge of the corporate context to science required a more “bottom-up” approach to conducting engaged sociotechnical field research. Our established expertise in developing metrics for estimating the health and sustainability of open source projects  ~\citep{germonprez_eight_2018, germonprez_rising_2019} helped initiate our engagement with open source scientific software projects. As newcomers, our interactions focused on questions about our expertise and how we could help~\citep{collins_rethinking_2008}, although we did not experience the level of skepticism toward expertise emerging in society~\citep{collins_rejecting_2014}.  

We think, in general, new field engagements where the researcher aims to engage over a long term rely on bridging knowledge of new, adjacent social and practice networks and are likely to begin as “bottom-up” rather than “top-down” relationships. However, the development and evolution of knowledge in adjacent social and practice networks for field researchers is not necessarily a linear or deterministic one where one can expect progression from expertise acquired from an adjacent network into impact and commitment in the new network.

The contrast between corporately engaged open source and open source scientific software illustrates the function of research context adjacency and distinction in the evolution of how researchers engage within and across networks and the role played by such engagement in advancing knowledge and community aims. Corporate open source software emerged during our long-term engagement as a context focused on increasing the systematic organizational structure of open source projects across its ecosystems for the strategic advantage of its members. The structure of the Linux Foundation, for example, is valuable to thousands of technology firms because it sustains essential infrastructure that is non-market-differentiating for its members. Moreover, maintaining this type of structured network of open source projects is actively facilitated by over a dozen annual conferences where all members or members with particular interests meet and build connections, ultimately serving the open source projects themselves and the Linux Foundation’s mission.   

Open source scientific software has two key characteristics that lead us to believe that effectively bridging adjacent social and practice networks is both a necessary and not routine research practice for sustaining a long-term research engagement. First, there are two distinct forms of scientific software interdependency among these open source projects, and they do not build from a singular foundation in the same way as the Linux Foundation centers around the most fundamental type of software: the operating system. Dependency type one includes observable, technical connections between open source scientific software projects. For example, many Matplotlib projects import another open source project for data manipulation called Pandas. Dependency types, not observable except through engaged field research in individual labs, are constructed from a combination of software and process within each lab’s “scientific pipeline.” In the life sciences, this includes a bricolage of strategies that move from some biological sample through data carpentry, analysis, and aggregation using distinct open source projects in concert with localized processes to perform each step required to produce research findings that advance research aimed at curing disease.  

A second key characteristic that guides us toward the sustained importance of bridging adjacent social and practice networks is the project leadership style in many open source scientific software projects and identifiable ecosystems. Leadership is more distributed~\citep{mcdonald_performance_2013, gronn_distributed_2000, gronn_leadership_2009, stewart_sustained_2015} than it is “lightly centralized”. We have not observed an innate or organic motivation to structure, survey, or organize the structural, technical, process, and social dependencies as potential leverage points in open source scientific software. In contrast, these factors significantly motivate corporate open source software contexts. Adapting knowledge to divergent materialities open source scientific software and corporate open source software are built from and aimed at different materialities, meaning that, methodologically, we must consider types of divergences to effectively adapt our knowledge to this new context~\citep{leonardi_materiality_2012}. In more philosophical terms, this form of embedded sociotechnical research compels an explicit recognition of how our ontological choices for framing the method may hide our epistemological commitments~\citep{goggins_connecting_2014, floridi_method_2008, floridi_web_2009} if our method is not clear about areas of divergence between sociotechnical contexts. For example, there is a divergence in the relationship between the social context of use for open source scientific software and corporate open source software. Small teams build open source software needed to do science in many domains, while corporate open source software coordinates, manages, and monitors across a much more extensive social network of developers and organizations.  

A methodology focused on long-term engagement and building knowledge in adjacent social and practice networks needs to foreground the importance of recognizing the entanglement of, in this case, the open source software, the social processes, and the research artifacts produced when interpreting the health and sustainability of open source scientific software~\cite{leonardi_materiality_2012,leonardi_whats_2010, barad_meeting_2007}. For example, the entanglement of people creating scientific and corporate open source software and the technologies themselves are different. It is more common in science for a small group of scientists to “build software they need,” even if that is not their primary job. In contrast, corporate open source software includes many software developers paid by corporations to build and maintain software their employers need.  

How we as researchers choose to make particular agential cuts~\citep{barad_meeting_2007} to distinguish the social, technical, and procedural objects we are interpreting will, effectively, result in a set of ontological commitments that define their relationship and constrain how we interpret observations~\citep{leonardi_materiality_2012, leonardi_whats_2010, barad_meeting_2007}. For example, one might distinguish these two contexts through the lens of motivation. That agential cut could be, “corporate open source is motivated by profit, while scientific open source is motivated by the production of papers.” Methodologically, it is not essential to make these “cuts” at the beginning, but developing the necessary knowledge of social and practice networks that are necessary to advance open source scientific software in the service of curing disease is already defining a critical agential cut we need to make to build and sustain relationships across domains during a long term engagement \citep{barad_meeting_2007}.  

\subsection{Context Adjacency \& Orthogonal Construct Operationalizations}

Our methodological care and attention to divergent materialities across networks are part of our method's reflexive nature. Metrics we defined through our work with CHAOSS and operationalized through our development of tools for gathering, analyzing, and triangulating electronic traces of the interactions between software objects and developers, between developers and developers, and between developers and projects fit other open source contexts well~\citep{goggins_group_2013}. Our knowledge of social constructs and software practices in this new open source context results from this prior work. 

Some sociotechnical research questions require long-term engagement, and in these cases, the researcher bears some responsibility for facilitating organizational structure in a new space adjacent to one they are already studying. In these cases, bringing value from the adjacent network and recognizing the new network have some critical differences, which are prerequisites to building a structured research program. In the case of open source \emph{scientific} software, its divergent materiality includes significant decentralization compared to corporate open source software. Understanding how and to what extent the current configurations in open source scientific software influence, advance, or impede scientific progress toward curing disease also demands the development of some social capital. We may begin building social capital by making invisible dependencies between open source software projects more visible or categorizing projects along dimensions this community thinks may distinguish a project essential or promising for accelerating the cure for a disease from those that do not. We may begin by operationalizing constructs from our work in corporate open source, which will help build knowledge of adjacent social and practice routines in three ways:

\begin{itemize}
    \item Illustrating metrics and analysis of open source scientific projects makes our potential usefulness more specific and clear.
    \item By operationalizing “health and sustainability” through metrics we built for a corporate environment, we identify needs, focus questions, and build social capital by using the rich data we have gathered to operationalize metrics that have more excellent utility in open source scientific software.
    \item Moreover, perhaps most critically, we will incorporate methods for identifying open source scientific software use by analyzing how and in what context scientific papers cite those projects.  
\end{itemize}

\section{Conclusion}

While researchers have deeply inspected individual methods for research, there is rarely a discussion of what is involved in establishing and continuing research agendas in studying large-scale sociotechnical systems. Thus, fundamental knowledge for research is relatively tacit or written off as relatively unimportant practice work.

This paper draws on our experiences and the literature on social capital to theorize and illustrate the work involved in long-term research agendas. We argue that understanding the distinct social and practice realms in an adjacent network is an underappreciated aspect of building research over long periods. Engaged field research in a first network leads to accomplishments, which makes understanding and appreciation of the adjacent network possible. Moreover, knowledge of the first network translates---albeit not without work---to utility in the adjacent context, shortening the time required to build relationships. This diligent work is somewhat invisible to evaluating academic productivity and almost absent from doctoral education, yet we feel it is knowledge worth knowing. 

Future sociotechnical scholars may leverage this methodological approach to advance their work into new worlds and contexts quickly. Alternatively, scholars may leverage this to explain why their particular approach to diligent work is vital for their research agendas, careers, and sociotechnical systems research. Finally, we hope others will apply their theoretical lenses to share, study, and clarify the long-term work involved in engaged scholarship.

{\label{687807}}

\selectlanguage{english}
\FloatBarrier

\bibliographystyle{apalike}
\bibliography{bibliography/references}

\pagebreak
\section{Appendix A}
\label{sec:example-section}
\begin{figure}[h!tbp]
	\centering
	\includegraphics[width=0.7\textwidth]{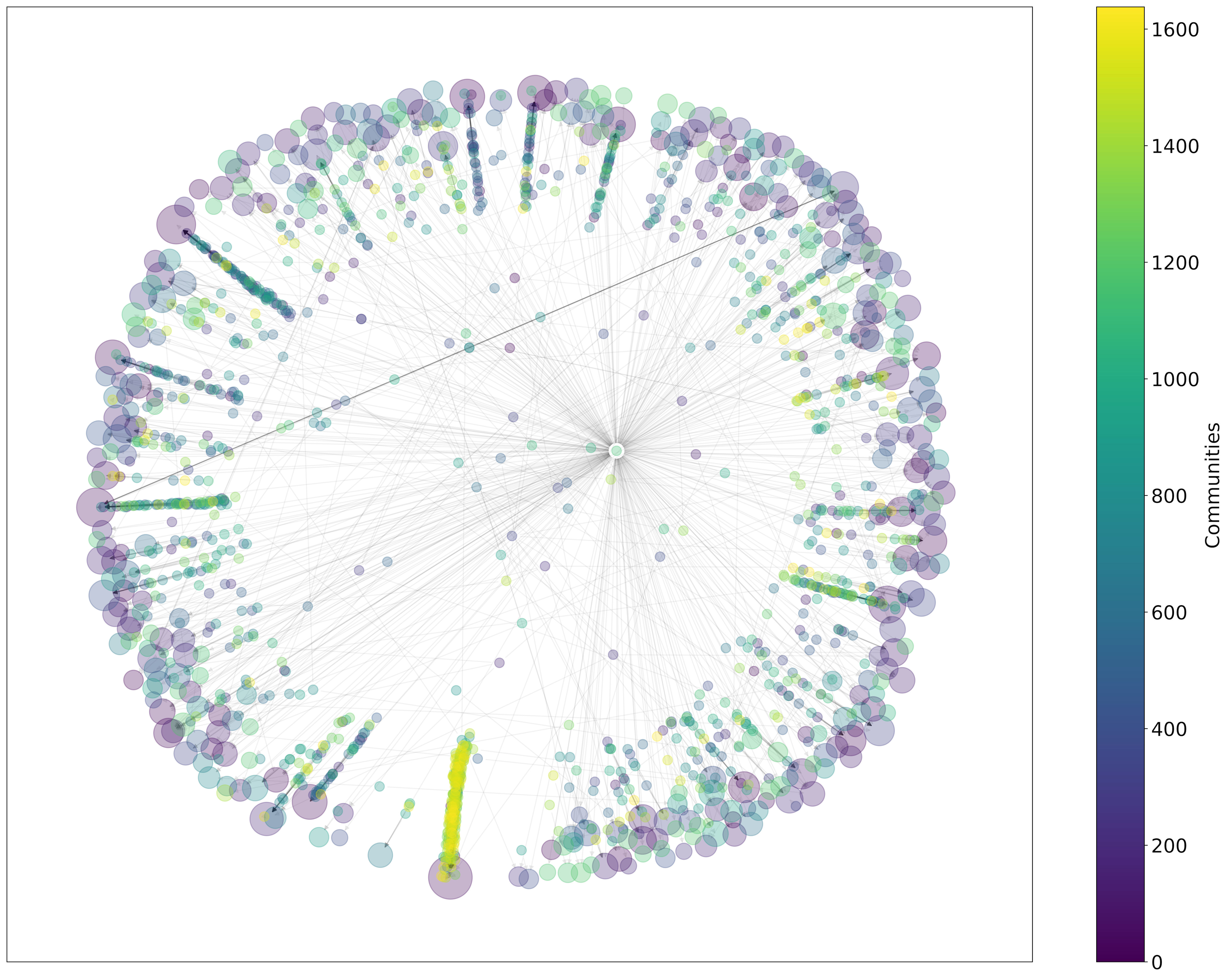}
	\caption{Dozens of these communities are visible in the donut-shaped corporate OSS ecosystem. }
	\label{fig:x corpsub}
\end{figure}

Through our collaborations with myriad scientific open source software groups, we identified and mapped comparisons between the corporate and scientific open source software structure.  Figures \ref{fig:x corpsub}, \ref{fig:x czi}, and \ref{fig:x harvardsub} in this appendix provide larger form visualizations of the summary level views in Figure \ref{fig:threegraphs}. Notable characteristics of these three sociograms include: 
\begin{enumerate}
    \item Network nodes are either contributors or repositories
    \item Repositories are purple
    \item Contributors are along the other color spectrum
    \item In Degree centrality is reflected in node size. Since only the repositories have In Degree centrality above 1 in our data organization, they are the only nodes that appear larger.
    \item  The distance between nodes in each graph is determined by the similarity of contributor patterns toward each of the repositories from the three ecosystems. 
    \item The shapes of each sociogram reflect the structure of each expert-identified ecosystem. 
    \item  We can see that in a large corporate ecosystem, the ecosystem's overall social structure is balanced across many projects. 
    \item In the two scientific ecosystems, there are defined centers, with funded scientific software showing several centers, and the organically identified ecosystem showing one center in a kind of funnel shape.
    \item These differences illustrate different practices and interconnections that define these ecosystems, succinctly illustrating that the social and contribution structures of scientific and corporate open source ecosystems are empirically different. 
\end{enumerate}

\begin{figure}[h!tbp]
	\centering
	\includegraphics[width=0.7\textwidth]{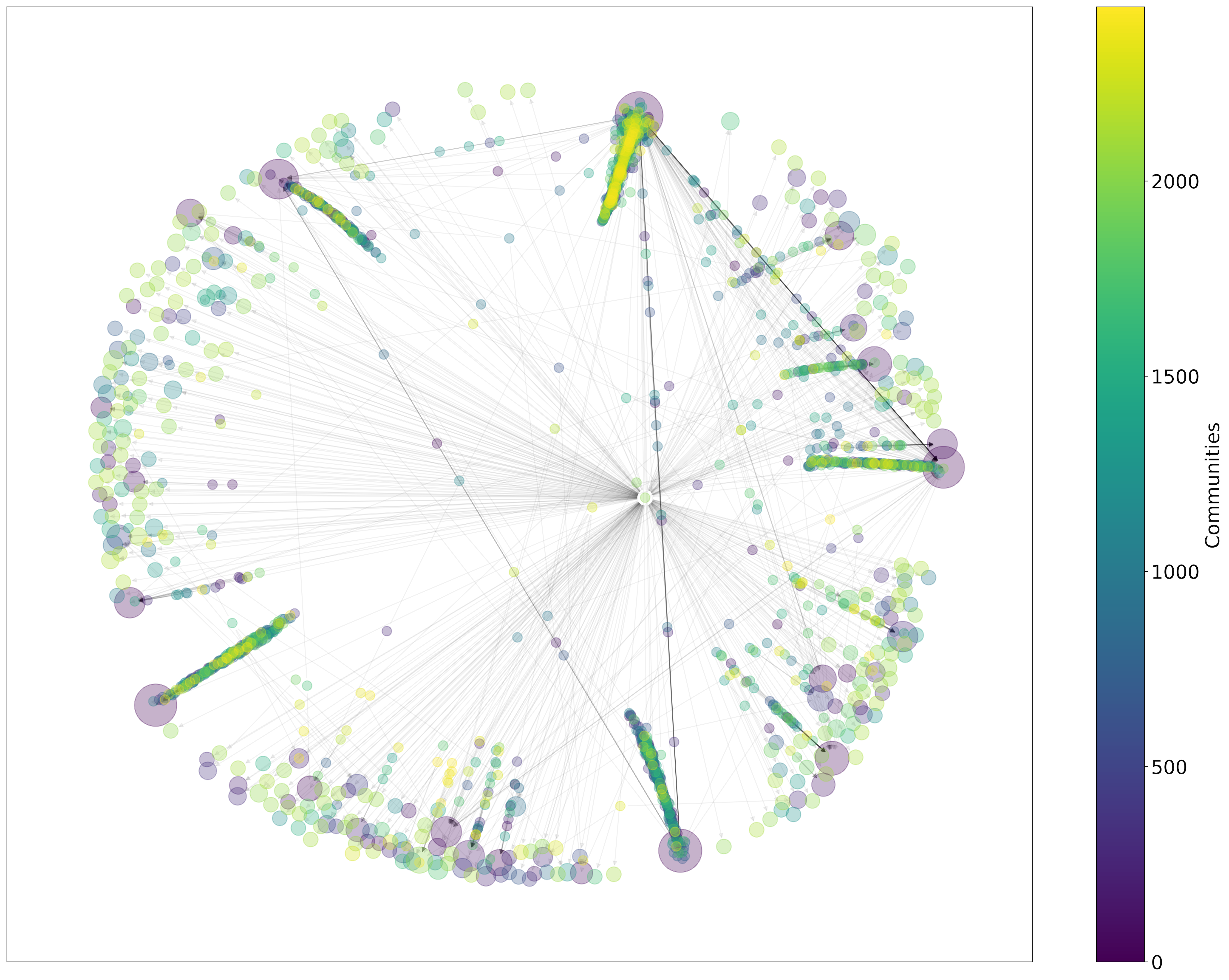}
	\caption{The clock-shaped funded scientific OSS ecosystem has 5-6 major community clusters.}
	\label{fig:x czi}
\end{figure}

\begin{figure}[h!tbp]
	\centering
	\includegraphics[width=0.7\textwidth]{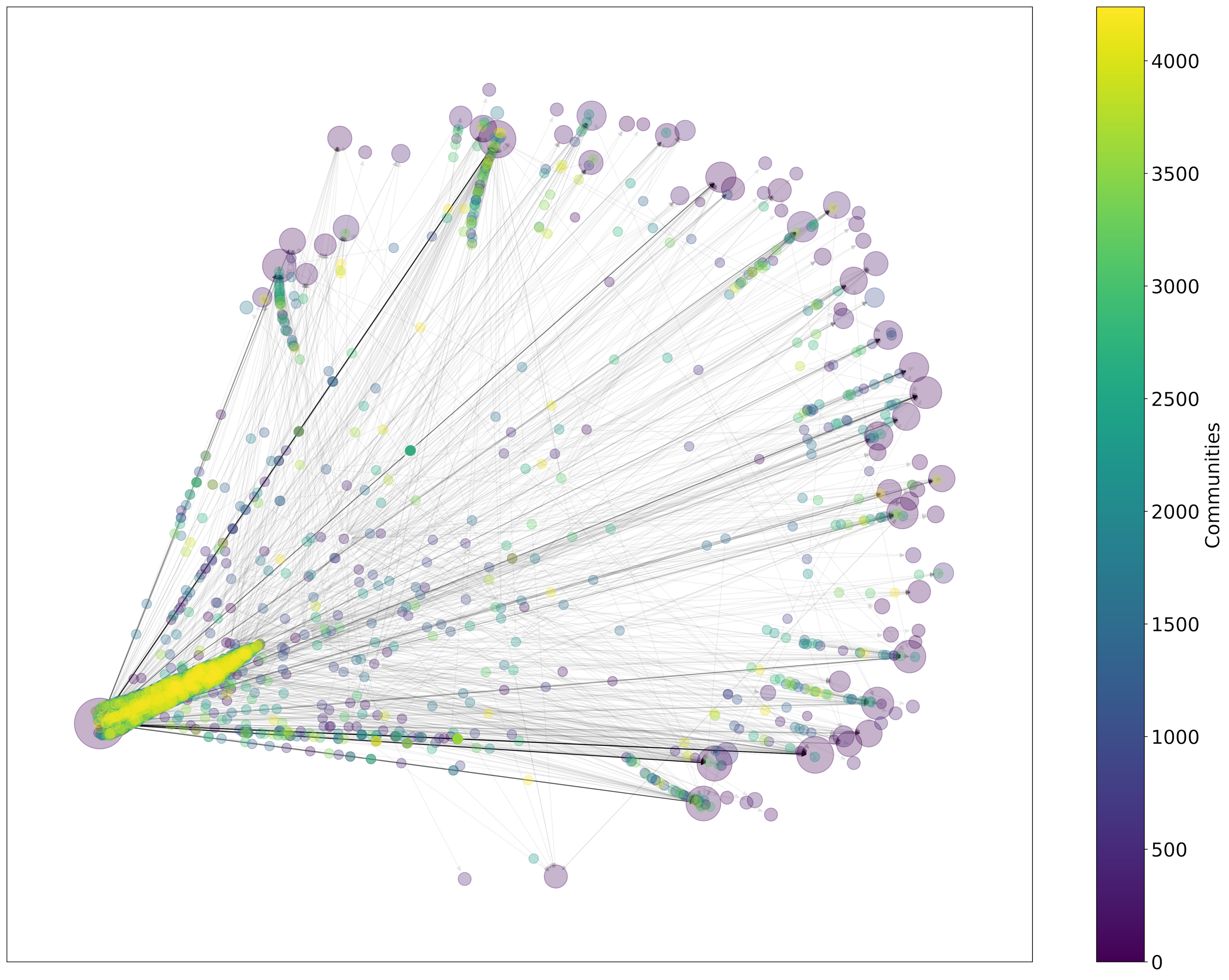}
	\caption{The funnel-shaped organic Scientific OSS ecosystem is visibly dominated by a singular community around a single, dominant project (in this case, Bioconductor).}
	\label{fig:x harvardsub}
\end{figure}

\end{document}